\documentclass[runningheads]{llncs}
\usepackage{makecell}
\usepackage[T1]{fontenc}
\usepackage{amsmath} 

\usepackage{graphicx}
\usepackage{multirow}
\usepackage[linesnumbered,ruled,vlined]{algorithm2e}
\usepackage{subcaption}
\usepackage{comment}
\usepackage[table]{xcolor}
\DeclareCaptionLabelFormat{andtable}{#1~#2  \&  \tablename~\thetable}

\begin{document}
\title{ARM SVE Unleashed: Performance and Insights Across HPC Applications on Nvidia Grace}
\titlerunning{ARM SVE in HPC}
%

\author{Ruimin Shi\inst{1} \orcidID{0009-0003-4387-367X}\and
Gabin Schieffer\inst{1} \orcidID{0009-0003-6504-7109}\and
Maya Gokhale\inst{2} \orcidID{0000-0003-4229-5735}\and
Pei-Hung Lin\inst{2} \orcidID{0000-0003-4977-814X}\and
Hiren Patel\inst{3} \orcidID{0000-0003-2750-4471}\and
Ivy Peng\inst{1} \orcidID{0000-0003-4158-3583}
}
\authorrunning{Ruimin Shi et al.}
%
\institute{KTH Royal Institute of Technology, Stockholm, Sweden \and
Lawrence Livermore National Laboratory, Livermore, USA \and
University of Waterloo, Waterloo, Canada}


\maketitle              

\begin{abstract}
Vector architectures are essential for boosting computing throughput. ARM provides SVE as the next-generation length-agnostic vector extension beyond traditional fixed-length SIMD. This work provides a first study of the maturity and readiness of exploiting ARM and SVE in HPC. Using selected performance hardware events on the ARM Grace processor and analytical models, we derive new metrics to quantify the effectiveness of exploiting SVE vectorization to reduce executed instructions and improve performance speedup. We further propose an adapted roofline model that combines vector length and data elements to identify potential performance bottlenecks. Finally, we propose a decision tree for classifying the SVE-boosted performance in applications.

\end{abstract}
\let\thefootnote\relax\footnotetext{pre-print submitted for publication}
\section{Introduction}
The landscape of processors on high-performance computing (HPC) systems has changed. For a long time, ARM processors have been dominating the embedded system market for their power efficiency, licensing flexibility, and wide toolchain support. In contrast, most HPC platforms have been powered by x86 processors, as represented by Intel and AMD processors. Although x86 processors are still used in many supercomputers, in the latest Top 500 list, 2 out of the Top 10 supercomputers in the world are powered by ARM processors, including the Supercomputers Fugaku~\cite{sato2021co} and Alps. Also, Jupiter, the upcoming exascale supercomputer in Europe, will be powered by ARM processors. This trend indicates that server-class ARM processors have emerged as a strong contender in high-end computing systems, especially due to increasing energy concerns, endorsement from vendors like Amazon and Nvidia, and diminishing gains from x86 processors~\cite{yokoyama2019survey,schieffer2024harnessing}.

Recent server-grade ARM processors have started to use the ARM scalable vector extension (SVE)~\cite{stephens2017arm} to increase computing throughput through vectorized instruction execution. Unlike conventional SIMD engines that only support a fixed vector length, SVE instructions can support a variable vector length by masking the predicate registers. This vector length agonistic (VLA) design is also used in the RISC-V vector extension (RVV)~\cite{asanovic2014instruction,lee2023test}. 
Unlike fixed-length SIMD designs, which require scalar loops to handle leftover tailing elements in irregular loops, and require instruction set expansion for every new vector length, SVE and RVV can support different vector lengths in one instruction set. 
Previous works~\cite{yamada2023optimization,wu2024autogemm,takahashi2023prototype} have optimized selected applications on specific ARM architectures. However, understanding the readiness and maturity to leverage ARM SVE in HPC applications remains an open question. 

This work aims to provide a first answer to the question. We propose a benchmark suite to reflect the evolving workload mixtures on HPC systems, including 13 applications from different domains, such as machine learning, drug discovery, scientific simulations, and quantum computing. This benchmark suite represents various code complexity, compute intensity and memory access patterns. We focus on compiler autovectorization in these applications as it is likely the most used approach for exploiting ARM SVE in existing applications. On a real hardware implementation of ARM SVE-- the Nvidia Grace processor--we leverage profiling and analytical models to evaluate how vectorized code can reduce the overall executed instructions and improve performance quantitatively.

We validate relevant performance hardware events on the ARM Grace processor and select a small set of events for deriving performance metrics. Though ARM processors have improved their support for PMU, there is still limited study of performance events on ARM processors. Based on the profiling results, we further propose an adapted roofline model that combines vector length in SVE architecture and data elements in applications to identify potential performance bottlenecks in applications. Guided by the roofline mode, we identified that both increased vector length and reduced data element sizes (via reduced data precision) could transform some compute-bound workloads into memory-bound workloads, highlighting the importance of matching memory subsystems on the vector architectures. Due to double-precision data formats, HPC applications cannot exploit short SVE like Grace CPU as much as single-precision machine learning workloads.

We propose a decision tree for classifying the performance speedup in applications on ARM processors with SVE into four classes. On 26 tested cases, 15 cases achieved speedup by ARM SVE without any porting efforts and 6 cases can be vectorized in compilation, have reduced retired instructions but cannot achieve performance speedup due to memory bound. In summary, we made the following contributions in this work.

\begin{itemize}
    \item We provide a benchmark suite of 13 applications to assess the maturity level of exploiting ARM SVE in HPC applications.
    \item We propose new metrics and validate hardware counters on Nvidia Grace processor for quantifying the vectorization effectiveness and identifying performance bottlenecks.
    \item We identify performance bottlenecks using an adapted roofline model combing vector lengths in SVE architecture and data element sizes in applications.
    \item We propose a decision tree for classifying the performance impact on ARM SVE and validate it in 26 cases on Nividia Grace processor.
\end{itemize}

\section{Background}

\begin{figure}[bt]
\centering
\begin{minipage}{0.48\linewidth}
\centering
    \includegraphics[width=\linewidth]{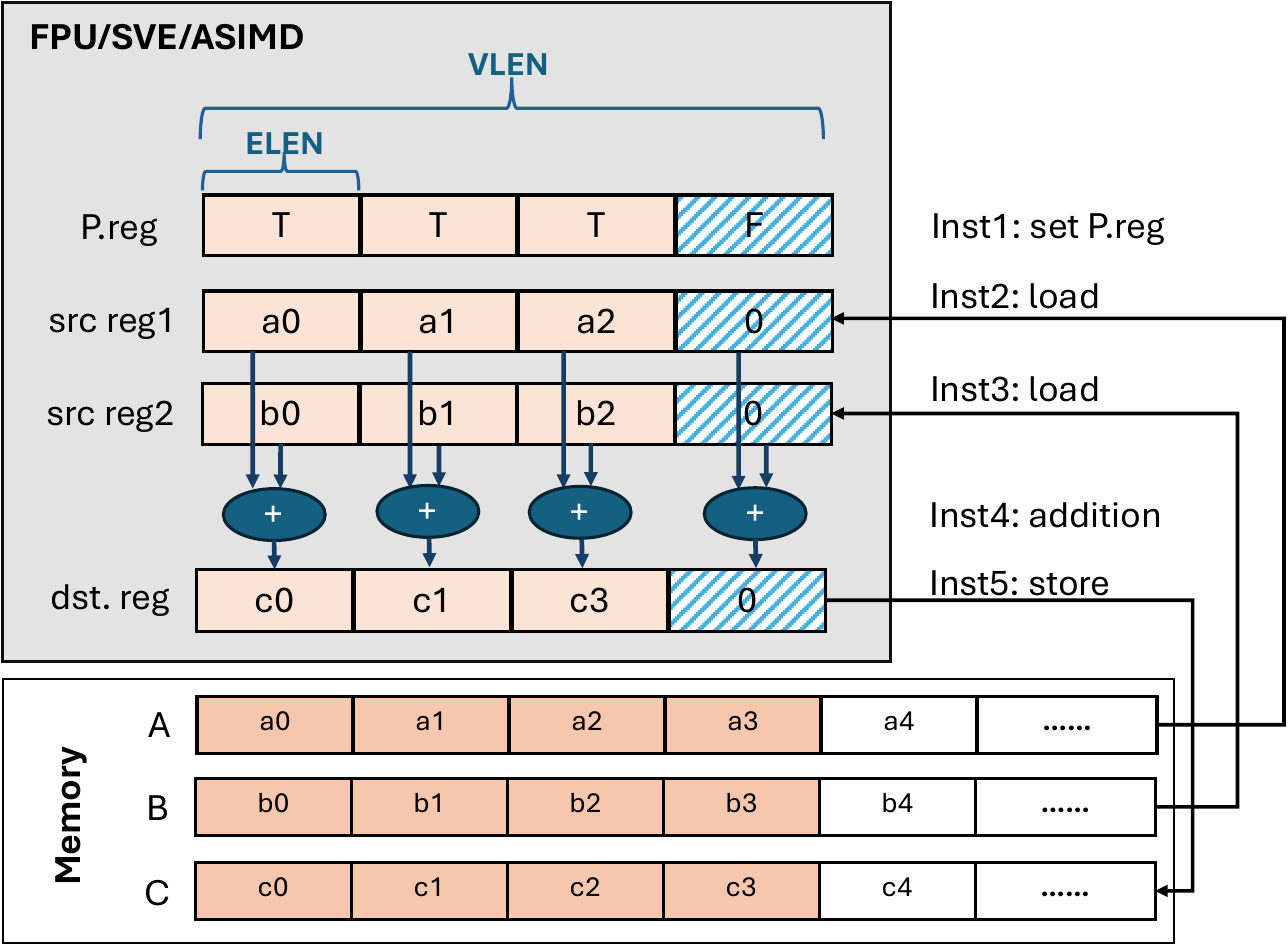}
    \caption{SVE supports variable vector length by masking predicate registers.}
    \label{fig:sve}
\end{minipage}\hspace{0.05\linewidth} 
\begin{minipage}{0.45\linewidth}
\centering
    \includegraphics[width=\linewidth]{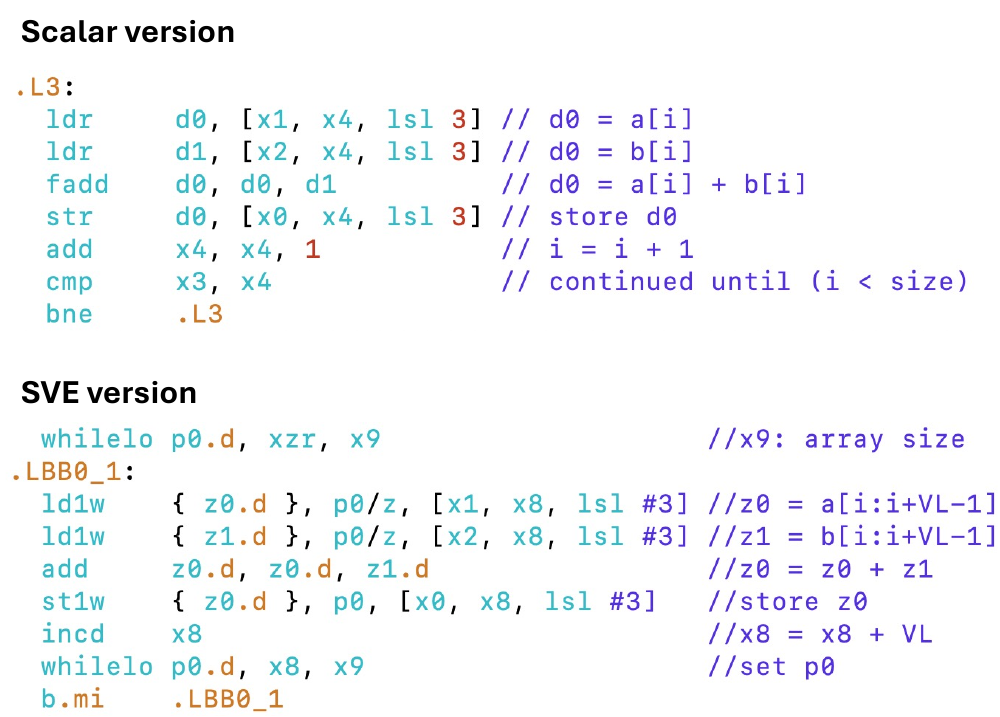}
    \caption{Assembly code of a simple vector-vector addition kernel.}
    \label{fig:vecadd}
\end{minipage}
\end{figure}

\textbf{ARM Scalable Vector Extension.}
Vector architectures explore data-level parallelism by simultaneously processing multiple elements in one instruction. Effective vectorization can reduce the number of retired instructions, improve the alignment of memory accesses, and increase computing throughput via parallel processing in arithmetic logic units (ALUs). ARM Scalable Vector Extension (SVE) is a long vector architecture introduced by the ARM A64 instruction set as part of the ARMv8-A and ARMv9-A architecture~\cite{saponarallama}. It has 32 vector registers and 16 predicate registers. Recent hardware implementations of ARM SVE include the Fujitsu A64FX processor (vector length 512-bit), Neoverse V1-based AWS Graviton processor (vector length 256-bit), and Neoverse V2-based Nvidia's Grace processor (vector length 128-bit).
ARM SVE exploit Vector Length Agnostic (VLA) programming models with vector registers ranging from 128-bit to 2048-bit at 128-bit increments on various architecture implementations. \texttt{VL} represents the number of elements operated by a specific instruction. Figure~\ref{fig:sve} illustrates a simple example where the fourth element is masked by setting predicate register (P.register) to false, and thus only three results are written back to the memory.  In this way, SVE achieves fine-grained control of each vector element through the setting of a predicate, enabling seamless handling of edge cases and exploring irregular data patterns. Figure~\ref{fig:vecadd} compares the scalar and SVE vectorized implementations of a simple kernel performing element-wise addition of two vectors.

\textbf{Software Support.}
To develop applications that exploit SVE engines, there are multiple approaches with different trade-offs between programming complexity and vectorization efficiency. Modern compilers support automatic vectorization that transforms a scalar code into a vectorized code using vector instructions. Auto-vectorization reduces programming complexity compared to writing SVE assembly, using SVE intrinsics, or relying on highly optimized ARM libraries.
By setting proper optimization flags, 
compilers can generate efficient vectorized codes without any rewriting. Auto-vectorization typically happens in loops, consecutive memory accesses, and tree data structures. For SVE, the GNU tools version 8.0+ and Arm Compiler (based on LLVM Clang) have vectorizers that detect suitable scalar operations to be optimized with SVE instructions. Auto-vectorization greatly reduces the programming overhead and improves portability across platforms. However, compiler-based auto-vectorization may suffer from insufficient optimization in complex codes. In this work, we focus on compiler-based auto-vectorization as the main approach for getting SVE adopted in realistic parallel codes, running on HPC systems.

\subsection{Related Works}



\textbf{Compiler Support.} 
Compiler support for vectorization have been widely studied. The vector effectiveness across the SVE-support compiler on mini-apps and SVE uasge are analyzed from instruction level~\cite{poenaru2020evaluating}. Source-to-source compiler-performed transformation has also been proposed as a solution~\cite{sato2023openacc,flynn2022exploring}. Specific auto-vectorization improvements have been proposed to compilers and runtimes~\cite{noor2022simd}. Instead of targeting application- or compiler-specific optimizations, our work assesses the opportunities and impact of compiler auto-vectorization for a variety of applications.


\noindent\textbf{Algorithm Co-design}. Several algorithms have been ported to use ARM SVE, focusing on providing insight on key design choices for vector architectures, either on simulators or actual hardware. These algorithms include the GEMM dense matrix-matrix multiplication routine~\cite{wu2024autogemm,wei2023dgemm} and Convolutional Neural Network (CNN)~\cite{gupta2023accelerating}. These works demonstrated the speedup of SVE vectorization for applications and presented different co-design methods and their effects. For guiding porting efforts, our work provides theoretical explanation, backed by a roofline model, to identify potential benefits of vectorization.


\noindent\textbf{Application Porting}. Porting efforts have been deployed to leverage ARM SVE in application codes, by using intrinsics. Examples include quantum simulator~\cite{takahashi2023prototype}, LLM model training~\cite{saponarallama}, DNA alignment tool~\cite{langarita2023porting}, and the widely-used NumPy Python library~\cite{yamada2023optimization}. We provide a method for application developers to identify potential benefits -- or non-benefits -- of auto-vectorization, based on their application's characteristics.


\section{Methodology}
\label{sec:method}
\textbf{Auto-vectorization} We investigate both GCC compiler and ARM compiler for autovectorizing HPC applications. Our results indicate that binaries vectorized by GCC compiler obtain better performance than the ARM compiler for most applications in test. Thus, if not specified otherwise, GCC compiler is used. We use three sets of compilation flags to create three versions for each application. First, the baseline version that only uses scalar instructions (denoted as Baseline) is obtained by disabling all vectorization options. In particular, we used \texttt{-fno-tree-vectorize} to disable vectorization on trees, \texttt{-fno-tree-loop-vectorize} to disable loop vectorization, and \texttt{-fno-tree-slp-vectorize} to disable the basic block vectorization. The second version uses the default Advanced SIMD (denoted as ASIMD) for vectorization, which is compiled with optimization flags \texttt{-march=armv8-a+simd} and 
\texttt{-mcpu=neoverse-v2}. Finally, a third version (denoted as SVE) is created by specifically enabling the SVE vectorization using compilation flags  \texttt{-march=armv8.5-a+sve} and \texttt{-mcpu=neoverse-v2}. We also verify the vectorization of generated codes by using \texttt{\$(CC) -S} to dump assembly code into \texttt{.s} files, then searching the identifiers of predicate and vector registers, such as \texttt{z0-z31}, \texttt{v0-v31} and \texttt{p0-p15}, to locate the vectorized code regions, and then confirm correct vector instructions are in use.

\textbf{Experimental Platform} We conduct our experiments on real hardware by using a testbed of the Nvidia Grace CPU~\cite{nvidia-grace-arch,schieffer2024harnessing}. The processor features 72 Armv9-A Neoverse V2 cores~\cite{armneoversev2}, equipped with 480~GB LPDDR main memory. It has L1 64KB I-cache and 64KB D-cache per core, 1MB L2 cache per core, and 117~MB LLC. The core also implements four 128-bit SIMD functional units, which are able to execute both SVE/SVE2 instructions and Advanced SIMD (also known as NEON) instructions. In this architecture, the maximum CPU frequency is 3447 MHz with four FPU pipelines per core, and the memory bandwidth tested by STREAM Triad benchmark is 30~GB/s and 250~GB/s at 1- and 72-threads, respectively. The system runs RHEL 9.4 with Linux kernel 5.14. GCC 11.4 compiler and ARM clang 23.10 are used on the platform. 

\subsection{Profiling Approaches}\label{sec:prof}

\begin{minipage}[b]{0.5\textwidth}%
\centering
\captionof{table}{ARM PMU events used for profiling on the ARM testbed.}
\label{tab:events}
\resizebox{\linewidth}{!}{
    \begin{tabular}{|c|c|c|c|c|}
    \hline
        Hexcode &Event Name &Description \\\hline
        0x8 & INST\_RETIRED &  Instruction executed  \\\hline
        0x37 &LL\_CACHE\_MISS\_RD &  LLC read, miss    \\\hline
        0x66  &MEM\_ACCESS\_RD& Memory access, load  \\\hline
        0x24 & STALL\_BACKEND&  Cycles due to backend stall\\\hline
        0x11 & CPU\_CYCLES  &   Cycles          \\\hline
         0x75 & VFP\_SPEC & Floating-point instruction\\\hline
    \end{tabular}
}
\end{minipage}%
\hspace{.5cm} 
\begin{minipage}[b]{0.4\textwidth}%
\centering
\captionof{table}{Benchmark suite}
\label{tab:benchmarks}
\resizebox{\linewidth}{!}{
    \begin{tabular}{|c|c|c|c}
    \hline
    \textbf{Application} &\textbf{Kernels} &\textbf{Problems} \\\hline\hline
LLM training  &train & 124M \\\hline
LLM inference &test  & 124M \\\hline
QC simulator  &RX\_gate & 21 qubits \\\hline
FFT1D         &fft1D    & 16384 \\\hline
FFT2D         &fft2D    & 262144 \\\hline
STREAM        &copy &1-10G \\\hline
DGEMM         &dgemm (FP64)& 12kx12k \\\hline
SGEMM         &sgemm (FP32) &12kx12k  \\\hline
SPMV          &spmv\_csr    &$2048^2$ \\\hline
Jacobi2D      &sweep &4-32k \\\hline
YOLOv3        &detector&  $608^2\times3$ \\\hline
AlexNet       &classifier &1k \\\hline
AutoDock      &scoring & \texttt{1iep}~complex \\\hline
\end{tabular}
}
\end{minipage}%
\vspace{10pt}

We extend a lightweight profiler library based on \verb|perf | to profile the instruction and memory details of selected kernels~\cite{perf,miksits2024multi}. With this profiler, we can collect the hardware counters provided by the ARM PMU in regions of interest (ROI). This profiler provides simple API in C/C++: \texttt{configure\_measure()} will configure and initialize the counters; \texttt{start\_measure()} and \texttt{stop\_measure()} will enable/resume and disable/pause counting for the hardware events; \newline 
 \texttt{print\_results()} will print the value of results into the terminal. 

The profiler uses the Linux system call \texttt{perf\_event\_open} to create a special file descriptor, each recording the measurement from an event. We group multiple descriptors together to set up different events at one time. 
In Neoverse V2 cores, at most six events can be collected simultaneously. 
The configuration structure of events uses \texttt{PERF\_TYPE\_RAW} where an event hexcode can be looked up for the specific hardware implementation. Table~\ref{tab:events} lists the events used in this work. For instance, we use retired instructions to quantify the vectorization effectiveness in the two vectorized versions, i.e., the ASIMD and SVE versions. Compared to static instructions that remain constant for the same executables, dynamic retired instructions reflect executed instructions on hardware. 

We validated a set of hardware counters provided by ARM PMU and found that some of them are not stable or accurate enough to be used in calculating evaluation metrics, like \texttt{STALL\_BACKEND\_MEM}, \texttt{L3D\_CACHE\_LMISS\_RD} and \texttt{SVE\_INST\_SPEC}.
Perf also provides \texttt{simd\_percentage} metric that is defined as the ratio between \texttt{ASE\_INST\_SPEC} and \texttt{INST\_SPEC} to represent the impact of vectorization. However, since both the numerator and denominator are capturing only speculatively executed instructions, 
\texttt{simd\_percentage} cannot reflect the overall reduction of total executed instructions. Instead, \texttt{INST\_RETIRED} is the architecturally executed instructions, which are more reliable without the interference of speculative execution in the superscalar processor.

\subsection{The Benchmark Suite}
\label{sec:apps}
We compose a diverse set of benchmarks that covers different application domains on HPC systems, code complexity levels, and various compute intensity and memory access patterns. We selected 13 applications from scientific simulation, machine learning, and quantum computing. 
We leverage our profiler to focus on key computational kernels and exclude initialization, preprocessing, and finalization stages, such as reading and preprocessing the input image in YOLOv3 and AlexNet, and loading real data matrices in SpMV. Table~\ref{tab:benchmarks} summarizes these applications with their respective largest input problems used, along with key computational kernels. These workloads support multi-threaded execution and we set the environment variables \texttt{OMP\_NUM\_THREADS} to control the thread count. Each experiment is repeated at least five times. We always guarantee the execution time of tests above 0.1s and the standard deviation within 5\%.

We also propose a synthetic benchmark based on SpMV ($y=Ax$) to configure different compute intensities and data formats. Assuming the sparse matrix A is stored in CSR format and \texttt{x} is the vector, each row is accessed by iterating through its nonzero elements, calculating \texttt{temp = val[j] * x[colind[j]] + temp}  to accumulate the results in spmv. 
Within one loop, three memory accesses are issued, and the load of \texttt{x[colind[j]]} is usually from the main memory due to its pointer-chasing nature with only two operations: \texttt{*} and \texttt{+}. To increase compute intensity, we repeat the computation for a configurable number of times, e.g., 20 in this example. To ensure compilers do not optimize or eliminate dead code automatically, this region uses \texttt{\#pragma unroll loop(1)} and disables the dead-code elimination (DCE) optimization flag. In the modified version, the corresponding values request of \texttt{colind}, \texttt{val} and \texttt{x} stays in L1D cache after the first computation. When the number of repeated computations increases, the benchmark can transform from memory-latency bound to compute-bound.

\subsection{Analytical Models}
We leverage analytical modeling to derive a set of theoretical metrics to guide the evaluation. First, we derive the theoretical upper bound of the vectorization ratio, based on the maximum vector length (VLEN) and data element formats (ELEN), as shown in the following equation for VB. To approximate ELEN in a kernel, we choose the dominant data formats, i.e., if the main computation is in double-precision, FP64 will be used.

\newcommand{\customsize}{\fontsize{8pt}{8pt}\selectfont}   

We then use the profiler and captured events to obtain the achieved overall instruction reduction. For this, we define a metric called instruction reduction ratio (denoted as $R_{ins\_reduction}$) to quantify the end-to-end effectiveness of exploiting vectorization for a given scientific problem. As shown in the equation below, it is defined as the ratio between the number of retired instructions using the non-vectorized version with the two vectorized versions using SVE and SIMD, respectively.
$R_{ins\_reduction}$ quantifies the reduction of total retired instructions for solving the same computation, i.e., instructions to a solution. This metric could reflect the impact of vectorized instructions in overall instructions. If vectorized instructions only compose a small fraction of all executed instructions, this application cannot effectively exploit vectorization for acceleration.

\begin{equation}
     Vectorization  \; Bound (VB) = \frac{VLEN}{ELEN} \quad \quad R_{ins\_reduction} = \frac{Ins_{nonvec}}{Ins_{simd | sve}}
\end{equation}

Finally, we adapt the roofline model~\cite{williams2009roofline} to capture the vectorization bound in a computation kernel. We leverage the adapted roofline model to identify performance bottlenecks. The inflection points on the roofline model for scalar and SVE, shorted respectively as IRR and IRV, are defined as
\begin{equation}
    AI_{IRR} = \frac{Peak \; Compute \; Throughput}{Peak BW} \quad \quad \quad 
    AI_{IRV} = AI_{IRR} * \frac{VLEN}{ELEN}
\end{equation}

PeakBW represents the peak achievable memory bandwidth on the platform.
If the arithmetic intensity (AI) of a kernel is smaller than the inflection point, this application is memory-bound, and increasing the memory bandwidth is the key optimization direction while vectorization cannot bring performance benefits. If its AI is greater than the inflection point, the application is compute-bound, and increasing the peak performance via vectorization can bring performance improvement. 
\section{Vectorization Effectiveness}
\begin{figure}[bt]
    \centering
    \begin{subfigure}{0.49\linewidth}
        \includegraphics[width=\linewidth,height=0.5\linewidth]{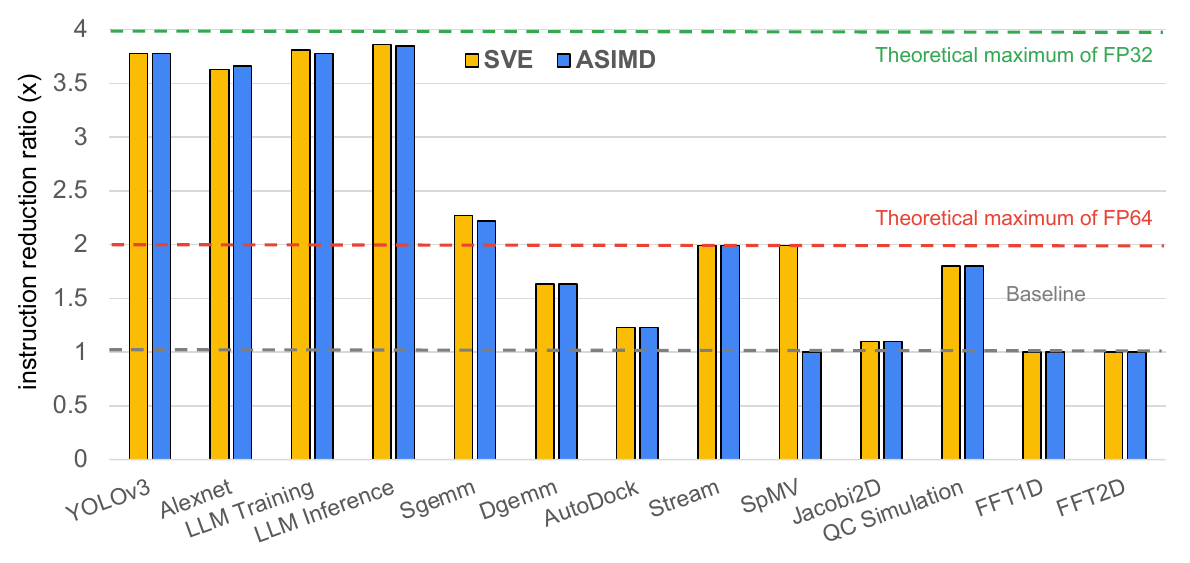}
        \caption{Instruction reduction ratio}
    \label{fig:vec-ratio}
    \end{subfigure}
\hfill
    \begin{subfigure}{0.49\linewidth}
        \includegraphics[width=\linewidth,height=0.5\linewidth]{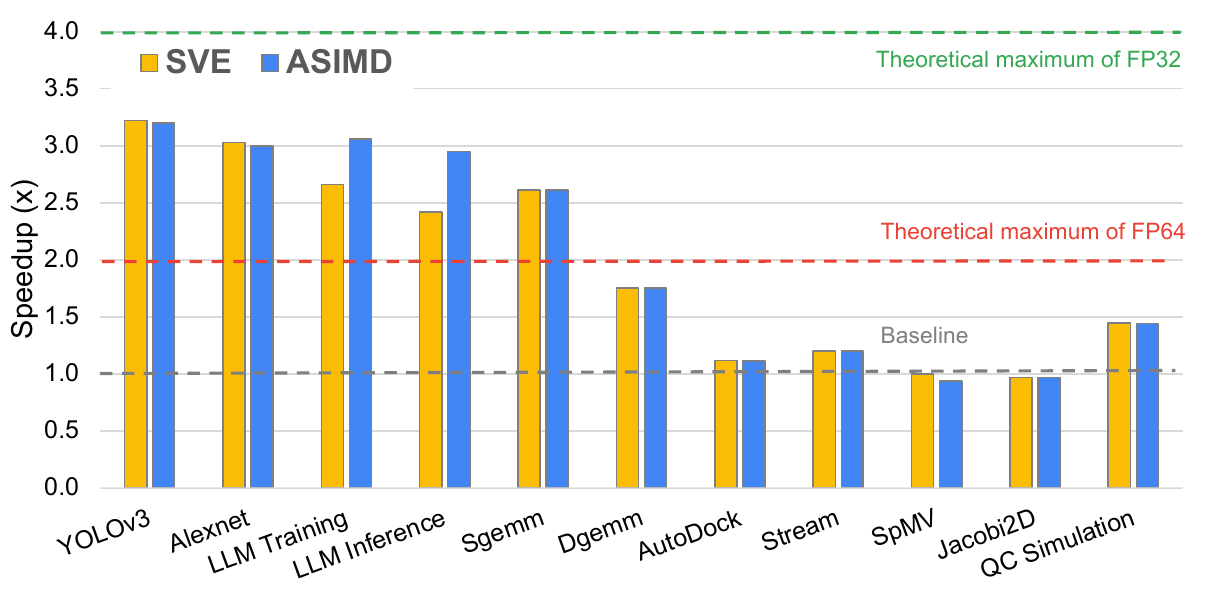}
        \caption{Achieved performance speedup }
    \label{fig:speedup-1t}
    \end{subfigure}
    \caption{Measured metrics in the 11 workloads that can be autovectorized using SVE and SIMD, respectively.}
\end{figure}

We present an overview of vectorization effectiveness in 13 applications by quantifying the reduced ratio of total executed instructions, i.e., $R_{ins\_reduction}$, in single thread execution in Figure~\ref{fig:vec-ratio}. On the Grace CPU, since the maximum vector length is 128~bits, the upper vectorization bounds (VB) for FP64 and FP32 data elements are 2$\times$ and 4$\times$, respectively, as indicated by the two dashed lines on Figure~\ref{fig:vec-ratio}. For each vectorized code, we also check its assembly code to confirm vectorization instructions are used. 11 out of 13 applications can be vectorized by compilers (via checking assembly code), and also have $R_{ins\_reduction}>1$. 
Many of the workloads achieved reduction ratios close to the vectorization bound. For instance, YOLOv3, AlexNet, LLM training, and LLM inference are single-precision workloads with $VB=4$ and they achieved 3.6-3.8$\times$ reduction in retired instructions. DGEMM, STREAM, and quantum circuit simulation are double-precision workloads with $VB=2$ and they achieved 1.6-1.8$\times$ reduction.

Only one application, the FFT benchmark in 1D and 2D, has limited auto-vectorization. This benchmark is implemented atop the FFTW subroutines library. By analyzing the source code and assembly code, we attribute this lack of auto-vectorization to complex intrinsic and pre-optimization based on the Radix-N algorithm in library design, requiring manual porting efforts for effective utilization of vector instructions. Note that the latest FFTW supports the ARM Neon extension, but still lacks support for SVE. 

SVE and Advanced SIMD have similar vectorization ratios for 12 benchmarks, except SpMV. The SpMV benchmark shows the advantage of SVE in processing the dynamic irregular loop length. In this benchmark, the loop lengths vary because the number of non-zero elements in each row of the sparse matrix is different, and this variability is challenging for the compiler to determine at compilation time before execution. Unlike the advanced SIMD, which relies on padding with scalar instructions on tailing elements or programmers' efforts to match the fixed vector length, SVE can use the predicate registers to manage the variable vector length at runtime. As a result, SVE achieved a $1.99\times$ instruction reduction ratio, whereas advanced SIMD only reaches $1.0\times$. 


We further compare the achieved performance speedup (Figure~\ref{fig:speedup-1t}) with the reduced instruction ratios (Figure~\ref{fig:vec-ratio}) for 11 applications in the benchmark suite, except the FFT benchmarks that cannot be vectorized. The four applications that have the highest reduction of retired instructions, i.e., YOLOv3, AlexNet, LLM training, and LLM inference, also achieved the highest performance speedup of $2.4-3.2\times$. However, the performance speedup in double-precision workloads is more diverse, DGEMM and QC simulation achieved $1.5-1.8\times$ speedup, which is consistent with their high $R_{ins_reduction}$. However, STREAM and SpMV show little performance improvement from vectorization, even though their retired instructions are effectively reduced by almost $2\times$.

Though most applications have the same speedup using SVE and ASIMD, LLM training and inference have a higher speedup with ASIMD than SVE. In LLM applications, the loop length in every layer is variable and parts of them have the dependency of previous results. The overhead of frequently setting the vector length using the dynamic vector length of SVE in runtime is not negligible. This may result in a lower speedup on SVE than ASIMD, which uses the fixed vector length.

The overall evaluation in auto-vectorization, reduction of instructions, and achieved performance in 13 applications indicates that on recent platforms like Grace, the maturity of compiler support for SVE is already comparable to the long-term maturity in support for the advanced SIMD. The support for single-precision workloads is better than double-precision workloads. Since HPC applications mostly use double-precision floating-point data formats, they may have a lower chance of speedup on ARM processors that implement short SVE, compared to other workloads.

\subsection{Impact of Thread Counts}
\label{sec:threads}
\begin{figure}[bt]
    \centering
    \begin{subfigure}{0.49\linewidth}
        \includegraphics[width=\linewidth]{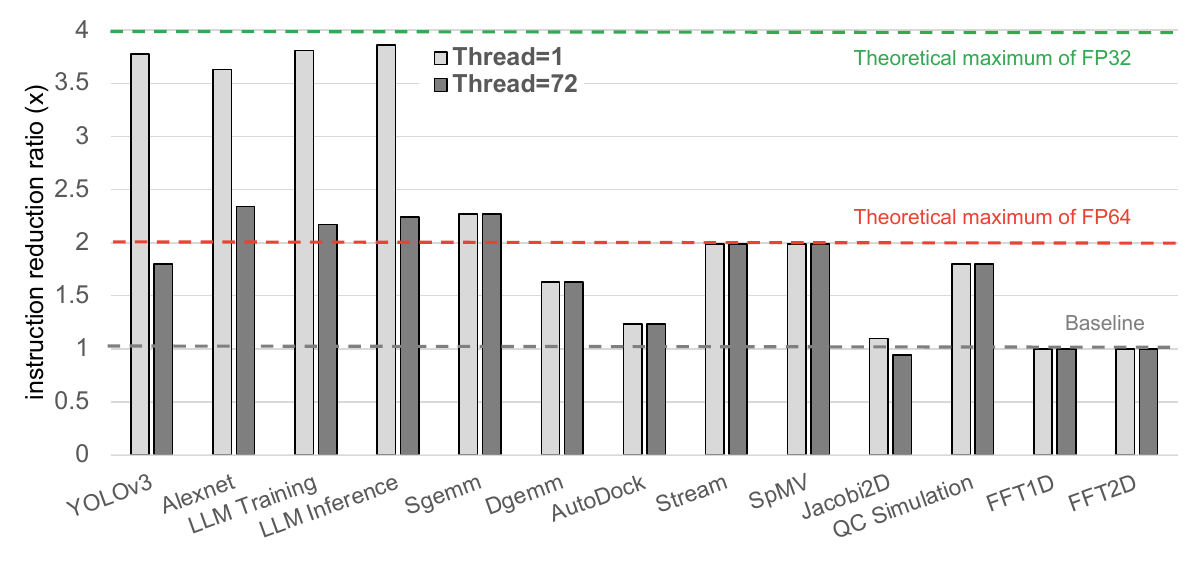}
        \caption{Instruction reduction ratio}
    \label{fig:thread}
    \end{subfigure}
    \hfill
    \begin{subfigure}{0.49\linewidth}
        \centering
        \includegraphics[width=\linewidth]{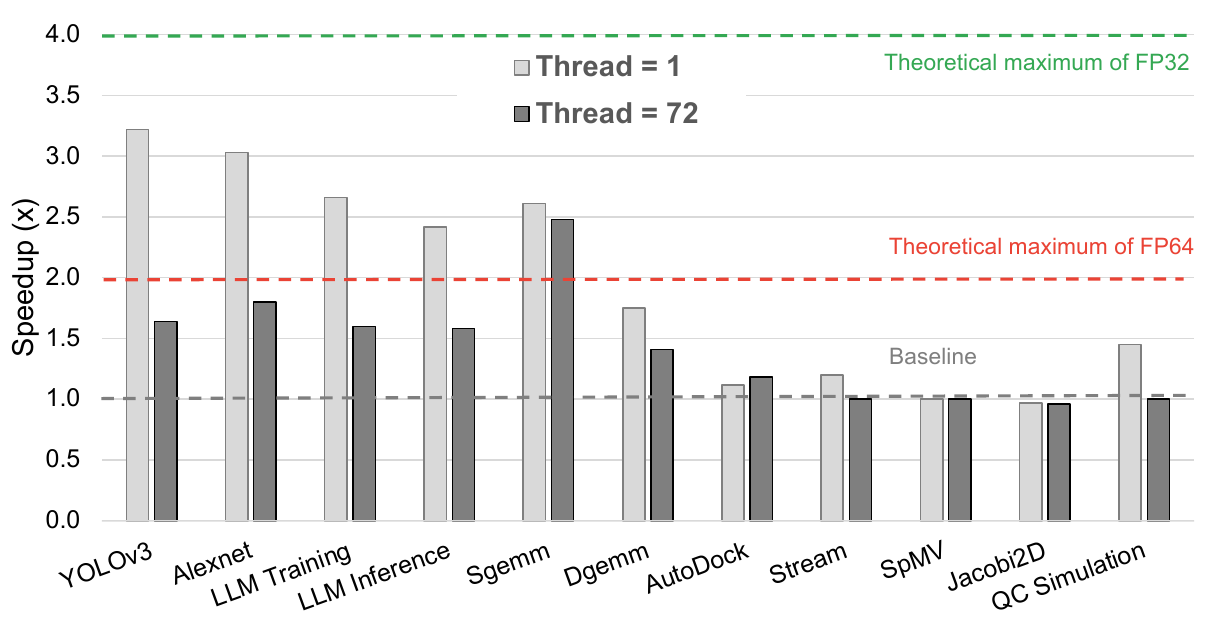}
        \caption{Achieved performance speedup }
        \label{fig:speedup-72t}
    \end{subfigure} 
    \caption{Measured metrics in 13 workloads using SVE on Nvidia Grace processor at single and 72 threads.}
\end{figure}

We compare the reduction ratios of total executed instructions at two thread counts in Figure~\ref{fig:thread}. Applications other than YOLOv3, AlexNet, and LLM training and inference maintain similar ratios at different threads, indicating that their total retired instructions are still reduced in the vectorized version when running with multiple threads. YOLOv3, AlexNet, and LLM applications are more complex applications with multiple tasks synchronized and control paths in parallelized loops. Their single-threaded runs achieve a good reduction ratio of retired instructions using vectorized versions. However, the reduction ratios at 72-threaded runs are much lower than those at a single-threaded run. One possible explanation could be the overhead in scheduling and synchronization on a large number of threads, when the codes call the dynamic OpenMP library to manage the multiple threads. The achieved performance speedup at different numbers of threads is presented in Figure~\ref{fig:speedup-72t}. As expected from their instruction reduction ratios, YOLOv3, AlexNet, LLM training and inference, show significant differences in speedup at the two thread counts.

\begin{figure}[bt]
    \centering
    \begin{minipage}{0.49\linewidth}
        \includegraphics[width=\linewidth, height=0.5\linewidth]{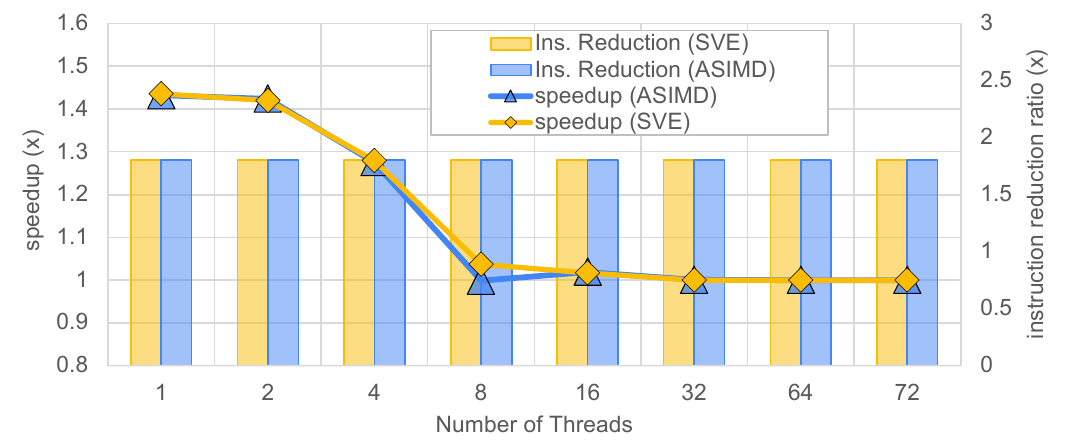}
        \caption{The speedup in the quantum circulation simulation.}
        \label{fig:speedup_QC}
    \end{minipage}\hspace{0.04\linewidth} 
    \begin{minipage}{0.46\linewidth}
        \centering
        \includegraphics[width=\linewidth, height=0.5\linewidth]{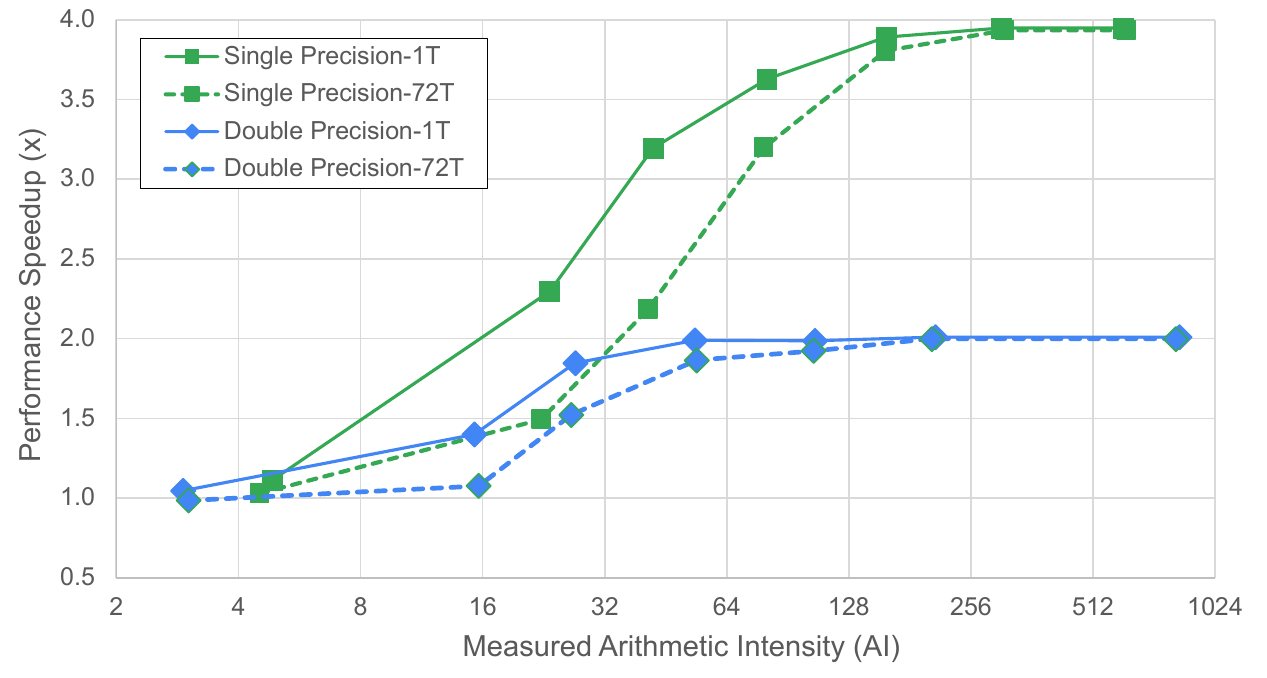}
        \caption{The speedup of the synthetic benchmark.}
        \label{fig:datatype}
    \end{minipage}
\end{figure}

Some applications, such as STREAM and quantum circuit simulations, show performance speedup from vectorization on single-threaded runs. However, with 72 threads, they no longer exhibit any speedup from the vectorized version even though the number of retired instructions is still effectively reduced in these runs. This behavior can likely be attributed to the increased memory contention when a large number of threads are running, potentially limiting the memory bandwidth and thus offsetting the potential performance speedup from using vectorization. Figure~\ref{fig:speedup_QC} presents the sensitivity test of a state-vector quantum circuit simulation of RX Gates at increased number of threads. Speedup of QC simulator decreases rapidly until the thread count equals 8, indicating the shift to memory-bound, and the saturation of memory bandwidth at 8 threads.


\subsection{Impact of Data Element Size}
We use the modified SpMV benchmark presented in Section~\ref{sec:apps} to assess the impact of element length on ARM SVE. We focus on SpMV as the overall performance results show that SVE is more flexible than SIMD for vectorizing irregular codes. Floating-point formats are ubiquitous in HPC applications, and on ARM architectures, three IEEE 754 compliant floating-point formats, i.e., double-precision (FP64), single-precision (FP32), and half-precision (FP16), are supported~\cite{armarchmanual}. Thus, we use these three data formats in the synthetic benchmark to evaluate 16-, 32-, to 64-bit element lengths. Given the 128-bit SVE vector length, each vector register should be able to hold 8, 4, and 2 data elements, respectively. However, half-precision (FP16) is excluded due to the lack of compiler support, i.e., GCC compiler throws compilation errors while ARM clang cannot generate vectorized code with correct \texttt{.h} instructions.

Figure~\ref{fig:datatype} presents the achieved speedup at different data element types and the measured arithmetic intensities. We changed the arithmetic intensity by increasing the computation repeat times, as described in Section~\ref{sec:apps}. For retired instructions, the reduction ratios are 2$\times$ in FP64 and 3.5$\times$ in FP16. It is clear from Figure~\ref{fig:datatype} that the performance speedup steadily increases as the compute intensity increases and saturates at around the calculated vectorization bound (VB), which equals 2 and 4 for FP64 and FP32, respectively. When the measured arithmetic intensity is low, their performance is bounded by memory, their achieved performance speedup is not proportional to the reduction ratio of retired instructions. 


To further investigate the compiler support for FP16, we also performed a test by changing the data types in the STREAM benchmark. We find that both GCC and ARM clang can correctly vectorize the code with \texttt{.h} instructions. Since the benchmark is bound by the memory bandwidth, no performance speedup is observed. However, the ratio of reduced instructions closely approximates the vectorization bound. When using GCC compiler, the instruction reduction ratio achieves $\times$2, $\times$4 and $\times$7.1 for FP64, FP32, and FP16, respectively; while ARM clang reaches $\times$2, $\times$4.4, $\times$3.5.

\section{Performance Modeling}
\label{sec:modeling}
We leverage a roofline model that combines SVE VLEN and applications ELEN to identify performance bottlenecks and further propose a decision tree that takes in profiling metrics from the non-vectorized version for classifying its performance impact on SVE-supported platforms.
\begin{figure}[bt]
    \centering
    \includegraphics[width=0.8\linewidth]{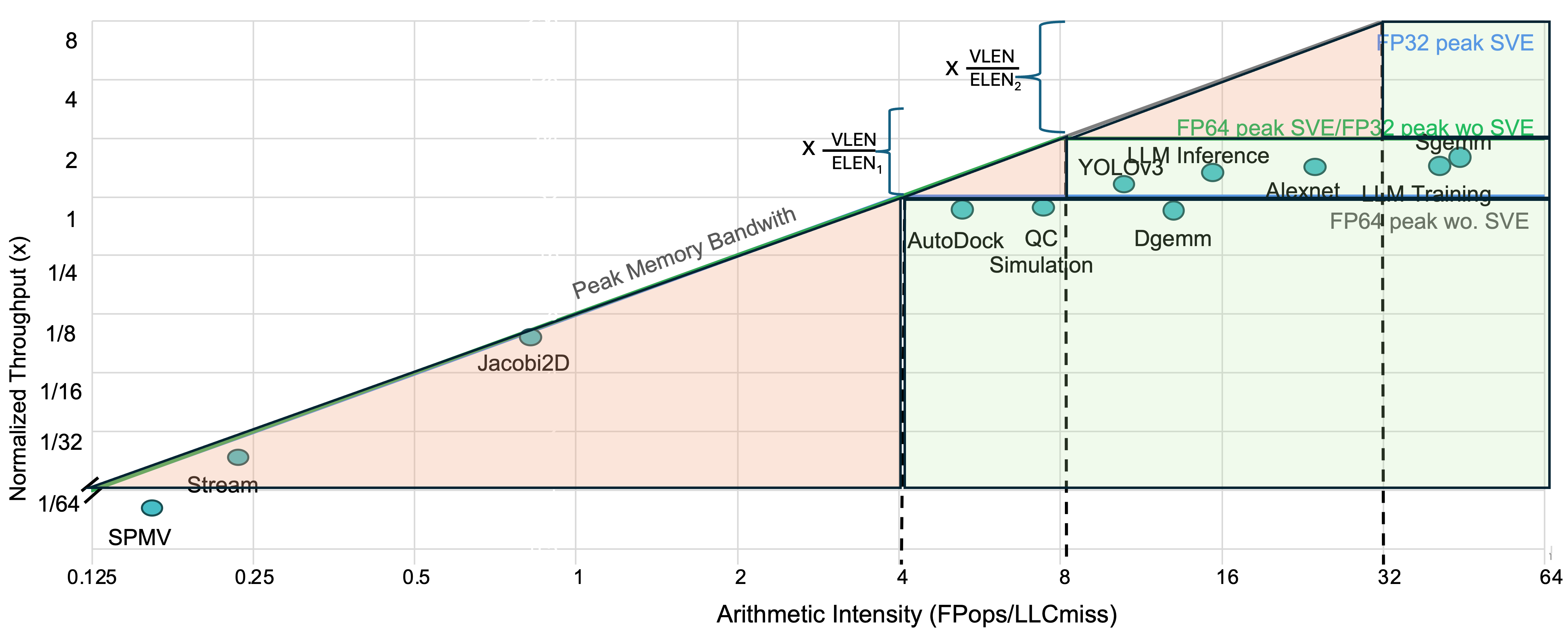}
    \caption{A Roofline Model that captures peak vectorization on a system with 128-bit SVE for single and double precision applications.}
    \label{fig:roofline}
\end{figure}

\textbf{Roofline model.} We propose a roofline model with extension for SVE architectures, as illustrated in Figure 9, to guide optimization directions in appli-
cations. We normalize the peak compute throughput with respect to the peak FP64 throughput without vectorization. In an ideal case, if an FP64 kernel is perfectly vectorized on 128-bit SVE, the peak compute throughput will be boosted by $2\times$. With an ideal vectorization in FP32, the peak compute throughput will be boosted by $4\times$ on 128-bit SVE. 
We validate the model by using the modified SpMV benchmark with increased computation repetition so that the benchmark moves from the memory-bound region to the compute-bound region. When using 20 repeated computations, the transformed kernel not only achieves $2\times$ reduction in retired instructions as its original version (in Figure~\ref{fig:vec-ratio}) but also achieves $1.8\times$ performance speedup unlike its original version (in Figure~\ref{fig:speedup-1t}), which has no performance speedup from vectorization. In fact, this achieved speedup is close to the $2\times$ boost in peak throughput in the roofline model. Furthermore, we change the data element size of the benchmark from FP64 to FP32. As predicted by the roofline model, when the ELEN is 32-bit, with 128-bit SVE, the modified benchmark achieves $4\times$ reduction in retired instructions and $3.8\times$ boost in peak throughput. In Figure~\ref{fig:roofline}, we annotate each application based on their estimated arithmetic intensity measured from single-threaded non-vectorized execution. The SpMV and STREAM benchmarks are in the memory-bound region (the left of the first dashed line). As their performance bottleneck is the peak memory throughput on the platform, vectorization cannot improve their performance. This is consistent with the observation that little performance speedup is reported in Figure~\ref{fig:speedup-1t} even though their retired instructions are reduced by vectorization in Figure~\ref{fig:vec-ratio}.
Theoretically, the vectorization can improve the peak performance corresponding to the architectural vector length. However, it may also move some workloads from compute-bound into memory-bound as illustrated in the two red triangles atop the green regions in Figure~\ref{fig:roofline}, like both quantum simulation and AutoDock. 
YOLOv3 and AlexNet are FP32 workloads and compute-bound. Thus, they both achieved performance improvement from SVE vectorization as reported in Figure~\ref{fig:speedup-1t}, and the vectorized version may also enter the red region. These observations indicate that pairing long vector engines with high-bandwidth memory may be important to avoid vectorized codes being bottlenecked by memory bandwidth. Such design is already used in Supercomputer Fugaku that uses HBM2 with 512-bit SVE~\cite{sato2021co}.

\begin{figure}[bt]
\begin{minipage}{0.49\linewidth}
    \includegraphics[width=\linewidth]{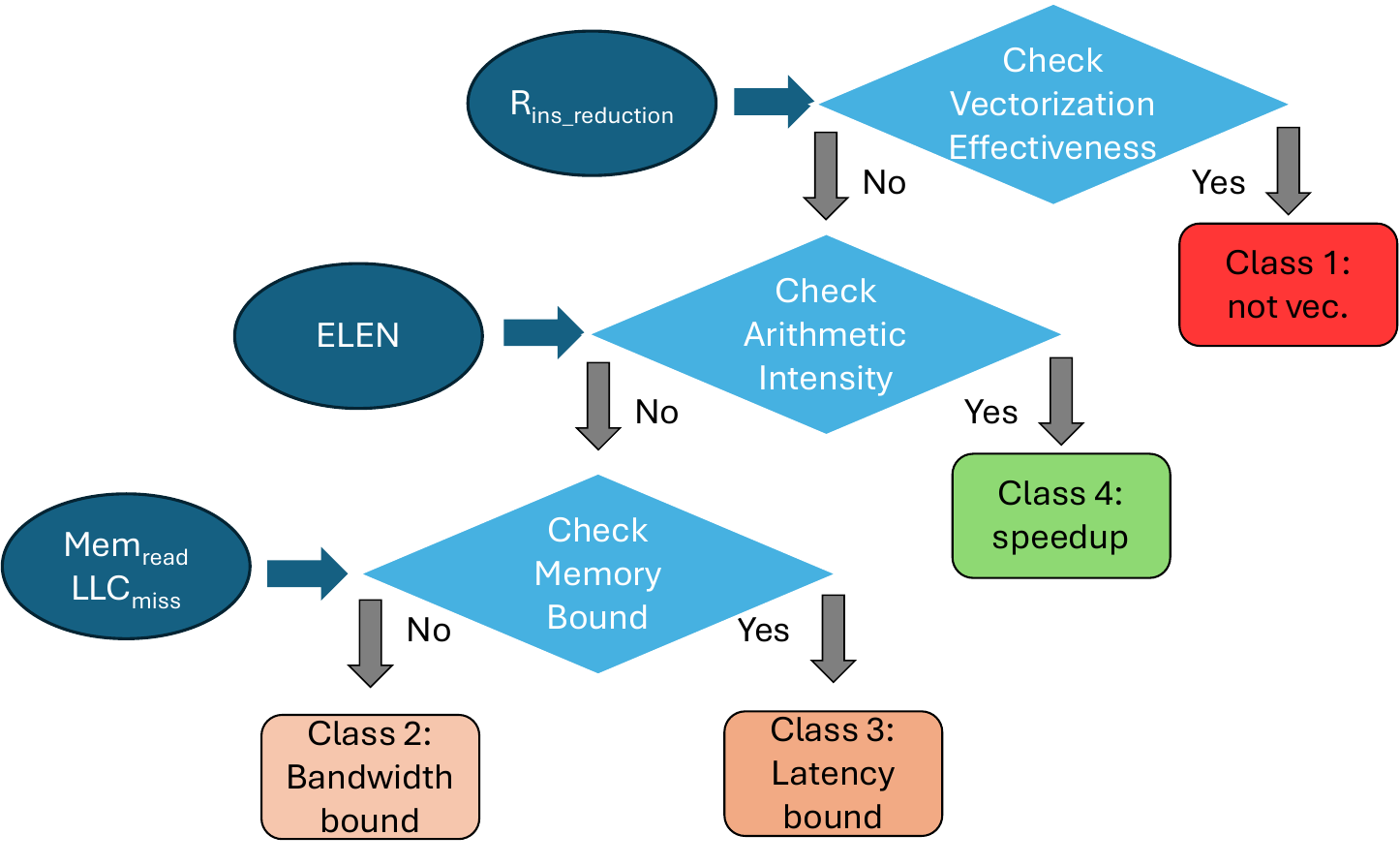}
    
\end{minipage} \quad \quad
\begin{minipage}{0.45\linewidth}
    \centering
\resizebox{\linewidth}{!}
{
    \begin{tabular}{|c|c|c|c|c|c|c|}
    \hline
\textbf{SN} &\textbf{Application} &\textbf{1-thread Case} &\textbf{72-thread Case}\\\hline
1   &YOLOv3         & \cellcolor{green!25}Class 4     & \cellcolor{green!25}Class 4\\\hline
2   &LLM training   & \cellcolor{green!25}Class 4     & \cellcolor{green!25}Class 4\\\hline
3   &LLM inference  & \cellcolor{green!25}Class 4     & \cellcolor{green!25}Class 4\\\hline
4   &QC simulator   & \cellcolor{green!25}Class 4     & \cellcolor{orange!25}Class 2\\\hline
5   &FFT1D          & \cellcolor{red!50}Class 1     & \cellcolor{red!50}Class 1\\\hline
6   &FFT2D          & \cellcolor{red!50}Class 1     & \cellcolor{red!50}Class 1\\\hline
7   &STREAM         & \cellcolor{orange!25}Class 2     & \cellcolor{orange!25}Class 2\\\hline
8   &DGEMM          & \cellcolor{green!25}Class 4     & \cellcolor{green!25}Class 4\\\hline
9   &SGEMM          & \cellcolor{green!25}Class 4     & \cellcolor{green!25}Class 4\\\hline
10   &SPMV          & \cellcolor{orange!50}Class 3     & \cellcolor{orange!50}Class 3\\\hline
11  &Jacobi2D       & \cellcolor{orange!25}Class 2     & \cellcolor{red!50}Class 1\\\hline
12  &AlexNet        & \cellcolor{green!25}Class 4     & \cellcolor{green!25}Class 4\\\hline
13  &AutoDock       & \cellcolor{green!25}Class 4     & \cellcolor{green!25}Class 4\\\hline
    \end{tabular}
}
    \label{tab:classification}
\end{minipage}
\captionlistentry[table]{A table beside a figure}
    \captionsetup{labelformat=andtable}
    \caption{Classification table(right) of the benchmark suite using the proposed decision tree(left).} \label{fig:tree}
\end{figure}

\textbf{Decision Tree.} Figure~\ref{fig:tree} presents the main required input and stages for determining an application as one of the four classes, i.e., not vectorized, memory latency bound, memory bandwidth bound, and speedup.
In this decision tree, we first determine whether a target kernel can be vectorized effectively using the introduced $R_{ins\_reduction}$ metric. This metric filters out those applications that either cannot be auto-vectorized by the compiler or their vectorized instructions only compose a small fraction of total executed instructions. The former scenario may be caused by the complex control logic inside the loops and the recursive calling of functions or libraries within the algorithm. The latter, as seen in Section~\ref{sec:threads}, could be caused by an increased number of non-vectorized instructions when other parts of the code such as threading runtime increase. The next step checks whether the kernel belongs to the memory- or compute-bound domains as shown in the roofline. For this, an estimated arithmetic intensity is used by approximation of $\frac{FP\_op}{LLC\_read\_miss}$. The LLC\_read\_miss is selected instead of the total memory accesses because memory accesses also included prefetching traffic while LLC\_read\_miss is more realistic to reflect the main part limiting the response speed from memory. The estimated arithmetic intensity is then compared against the inflection point $AI_{inflection}$ on the roofline model in Figure \ref{fig:roofline}. The memory-bound class is further divided into bandwidth or latency bound, by comparing the obtained last level cache miss ratio ($R_{llc}$) with an ideal miss ratio ($R_{llc}=\frac{ ELEN }{cache\_line}$), assuming an application with large input problem filling the LLC. On Grace processor, the LLC line size is 64-byte, with 8-byte double float-point, 13\% is used as the threshold value.

Using the decision tree, we managed to classify 26 cases from 13 applications in the benchmark suite. As summarized in Table 
3, 15 out of 26 tested workloads can be sped up by vectorization without any porting efforts. Five cases, including FFT1D and 2D, and Jacobi2D in 72 threads, cannot be vectorized effectively and thus no performance gain is obtained from SVE. Finally, four cases, including the quantum circuit simulation at 72 threads, STREAM, and Jacobi2d at single thread, can be vectorized by compilation, but they cannot achieve performance speedup on the platform due to memory bound.
\section{Conclusion}
This work assesses the maturity of exploiting ARM SVE in HPC applications. We provided a suite of 13 applications from machine learning, scientific, and quantum computing. We defined quantitative metrics, validated hardware counters, and proposed a roofline model for identifying performance bottlenecks on ARM SVE on the Nvidia Grace processor. With a decision tree, we managed to classify 26 cases and found that 15 achieved speedup via SVE vectorization. Our results indicate that double-precision HPC applications face more challenges in exploiting SVE than single-precision machine learning workloads because they need longer vector lengths and higher memory bandwidth.

\section*{Acknowledgment}
This research is supported by the European Commission under the Horizon project OpenCUBE (101092984). This work was supported by the Lawrence Livermore National Laboratory LDRD Program under 25-ERD-016. LLNL-CONF-2005986.

\bibliographystyle{splncs04}
\bibliography{main}
%





\end{document}